\documentclass[10pt,letterpaper,dvips]{article}
\usepackage{opex3}
\usepackage{amsmath}
\usepackage{graphicx}
\usepackage{subfigure}
\usepackage{epic}

\begin{document}

\title{Theoretical Study of the Spurious-Free Dynamic Range of a Tunable Delay Line based on Slow Light in SOA}

\author{Perrine Berger$^{1,2}$, J\'er\^ome Bourderionnet$^1$, Mehdi Alouini$^{1,3}$, \linebreak Fabien Bretenaker$^2$, and Daniel Dolfi$^1$}

\address{$^1$Thales Research \& Technology, 1 av. Augustin Fresnel, 91767 Palaiseau Cedex, France}
\address{$^2$Laboratoire Aim\'e Cotton, CNRS-Universit\'e Paris Sud 11, Campus d'Orsay, 91405 Orsay Cedex, France}
\address{$^3$Institut de Physique de Rennes, UMR CNRS 6251, Campus de Beaulieu, 35042 Rennes Cedex, France}

\email{Perrine.Berger@thalesgroup.com} 



\begin{abstract} 
We developed a predictive model describing harmonic generation and intermodulation distortions in semiconductor optical amplifiers (SOAs). This model takes into account the variations of the saturation parameters along the propagation axis inside the SOA, and uses a rigorous expression of the gain oscillations harmonics. We derived the spurious-free dynamic range (SFDR) of a slow light delay line based on coherent population oscillation (CPO) effects, in a frequency range covering radar applications (from $40\, \mathrm{kHz}$ up to $30\, \mathrm{GHz}$), and for a large range of injected currents. The influence of the high order distortions in the input microwave spectrum is discussed, and in particular, an interpretation of the SFDR improvement of a Mach-Zehnder modulator by CPOs effects in a SOA is given.  
\end{abstract}



\section{Introduction}\label{s1}
\noindent 
A continuously tunable optical delay line is a key element for a large variety of microwave photonics applications, including the control of optically fed phased array antennas, the filtering of microwave signals or the synchronization of optoelectronic oscillators \cite{Yao2009,Dolfi1996,Capmany2006}. Slow light based tunable delay lines have been therefore intensively studied over the past few years. The main focus has been to understand and modelize the underlying phenomena \cite{Gehring2008}, in order to increase the delays and/or the bandwidth of the studied effect \cite{Chen2008,Thevenaz2006}. Among the different slow and fast light architectures, coherent population oscillations (CPOs) in a Semiconductor Optical Amplifier (SOA) offer attractive operational advantages in terms of compactness, integrability and possible parallelism, as well as a continuous tunability of the delay, or phase shift, through the injected current. Moreover, a recent demonstration of $360\,^{\circ}$ phase shift obtained at up to 19 GHz using this technology, makes it very promising towards integration in radar systems.\\ 
The principle of slow light in SOAs can be understood as follows. An optical carrier with a sinusoidal modulation envelope that propagates in a SOA induces a carrier population oscillation, via gain saturation in the semiconductor material. Since gain oscillations are in antiphase with the modulation envelope, and basically occur for modulation frequencies below the inverse of the carrier lifetime, they induce a dip in the RF gain spectrum of the SOA. Consequently, according to the Kramers-Kronig relations, the gain dip is associated with a large and positive refractive index dispersion, and hence a large group index, thus slowing down the sinusoidal modulation propagation velocity. This property, as well as corresponding phase and amplitude change of the modulation envelope, have been extensively studied, both experimentally or theoretically \cite{Agrawal1988,Chang-Hasnain2006,Su2006,Mork2005}. In particular, this has provided the RF transfer function of the SOA-based delay line, similarly to any two-port electronic device.\\ However, within the scope of an integration in a realistic radar system, it is also required to study the impact of these slow light architectures on the whole microwave photonics link, and in particular the generation of spurious signals by harmonic and intermodulation products. 
One of the key parameters of a microwave photonics link for radar applications is indeed the spurious-free dynamic range (SFDR) \cite{Zmuda1994}. It represents the operational dynamic range of the link, i.e., the range of input (or output) RF powers, at a given frequency, for which all spurious signals at harmonic and intermodulation frequencies are below the noise level at the output of the link. The SFDR requirements for radar microwave photonics links are more or less stringent depending on the application, but are basically in the range of 105dB/Hz$^{2/3}$ (ground based antennas) to 115dB/Hz$^{2/3}$ (airborne antennas).   
Most of the previously reported work on nonlinear frequency mixing in SOAs has been carried out in the frame of Radio-over-Fiber applications, or SOA-based in-line photodetection \cite{Seo2003,Sharaiha1998,Udvary2003}. The experimental and operational context of these works significantly differs from radar applications. As an example, in \cite{Seo2003}, the SOA is used for multifrequency signal up-conversion, where incoherent frequency channels are considered. However, a typical radar microwave photonics link involves a single electro-optic modulator that indistinctly transfers all the spectral components of the received RF signal (including a jammer, for instance) to the optical carrier. Coherent beating between spectral components can thus not be avoided, and the SFDR measurements reported in \cite{Seo2003} can therefore not be extrapolated to slow light based tunable delay lines for radar systems. \\
In this paper, we first present a predictive model to determine the harmonic generation level, which takes into account the spatial evolution of the saturation parameters along the SOA length. We also discuss the expression of the gain modulation harmonics that is used in common models, and show that it can lead, under certain conditions, to a significant error on the harmonic's estimate. Then, we generalize our model to end up with the third order intermodulation distortion ($\mathrm{IMD_3}$), and compute the $\mathrm{IMD_3}$ level and the SFDR over the frequency range of interest for radar applications (up to $20$GHz), and for a large range of injected currents. Finally, we discuss the influence of the input optical spectrum on the dynamic range of the delay line by comparing the simulation results obtained for an ideal spectrum, i.e., a perfect sinusoidal modulation of the optical beam at the input of the SOA, and for a realistic spectrum, namely, when the RF signals to be delayed are transferred on the optical carrier using a standard Mach-Zehnder modulator.

\section{Principle of harmonics calculation}\label{s2} 

We consider an optical field $E(z,t)$ which propagates along a traveling wave semiconductor optical amplifier (SOA). The interaction of light with carriers in the SOA is governed by the well known carrier rate equation and field propagation equation \cite{Agrawal1988}:
\begin{eqnarray}
\frac{dN(z,t)}{dt}&=&\frac{I}{qV}-\frac{N(z,t)}{\tau_s(z)}-\frac{g(z,t)\left|E(z,t)\right|^2}{\hbar\omega}, \label{eq1}\\
\frac{d(|E(z,t)|^2)}{dz}&=&\big(-\gamma_i+\Gamma g(z,t)\big)|E(z,t)|^2 \label{eq2}
\end{eqnarray} 
where $N(z,t)$ is the carrier density, $I$ the current injected in the SOA, $\tau_s$ the carrier lifetime, $q$ the elementary electric charge, $V$ the SOA active volume, $g$ the optical gain, $\Gamma$ the confinement factor, and $\gamma_i$ stands for the internal losses.
In order to find the level of the generated harmonics, we first consider that the input optical field is modulated at the RF frequency $\Omega$. $|E|^2$, $g$ and $N$ are hence all time-periodic functions with a fundamental frequency of $\Omega$. They can therefore be written into Fourier harmonic decompositions:
\begin{eqnarray}
|E(z,t)|^2&=&\sum_{k=-\infty}^{+\infty} M_k(z) e^{-ik\Omega t}, \label{eq3}\\
N(z,t)&=&\bar{N}(z) + \sum_{\substack{k=-\infty \\ k\neq0}}^{+\infty} N_k(z) e^{-ik\Omega t}, \label{eq4}\\
g(z,t)&=&\bar{g}(z) + a(z)\sum_{\substack{k=-\infty \\ k\neq0}}^{+\infty} N_k(z) e^{-ik\Omega t} \label{eq5}
\end{eqnarray}
where $\bar{N}(z)$ and $\bar{g}(z)$ respectively denote the DC components of the carrier density and of the optical gain. $a(z)$ is the SOA differential gain, defined as $a(z)=\partial\bar{g}/\partial\bar{N}$. Defining $g_k$ as the oscillating component of the gain at frequency $k\Omega$, and considering only a finite number $K$ of harmonics, the carrier rate equation (\ref{eq1}) can be written in an equivalent matrix formulation:
\begin{equation}
\begin{picture}(350,120)
\thicklines
\dottedline{5}(135,100)(180,73)
\dottedline{5}(215,50)(260,23)
\thinlines
\drawline(135,50)(180,23)
\drawline(215,100)(260,73)
\drawline(135,85)(155,73)
\drawline(160,100)(180,88)
\drawline(215,35)(235,23)
\drawline(240,50)(260,38)
\put(0,60){$\left( 
\begin{array}{c}
0\\
\vdots \\
0\\
\hbar\omega \left(\frac{I}{qV}-\frac{\bar{N}}{\tau_s}\right)\\
0\\
\vdots \\
0
\end{array}\right)=
\left( \begin{array}{ccccccc}
\alpha_{-K} & M_{-1} & \hdots & M_{-K} & 0 & & 0 \\
M_{1} & & & \vdots & & & \\
\vdots & &  & M_{-1} & & & 0 \\
M_{K} & \hdots & M_{1} & \alpha_0 & M_{-1} & \ldots & M_{-K} \\
0 & & & M_1 & & & \vdots \\
 & & & \vdots & & & M_{-1} \\
0 & & 0 & M_{K} & \ldots & M_1 & \alpha_{K} 
\end{array}\right)\times
\left( \begin{array}{c}
g_{-K}\\
\vdots \\
g_{-1}\\
\bar{g}\\
g_1\\
\vdots \\
g_K
\end{array}\right)$}
\end{picture}
\label{eq6}
\end{equation}

where $\alpha_k=I_s(1+M_0/I_s-ik\Omega\tau_s)$, and $\alpha_0=M_0$ is the DC optical intensity. $I_s$ denotes the saturation intensity and is defined as $I_s=\hbar\omega/a\tau_s$. It is worth mentioning that $\alpha_k$ is obtained at the first order of equation (\ref{eq1}), when mixing terms are not considered. One can also notice that $g_{k}^*=g_{-k}$. For the sake of clarity, we did not write the dependence on the propagation coordinate $z$ in (\ref{eq6}). It is however important to note that in the following, $\bar{N}$, $\bar{g}$, $a$, $\tau_s$, $I_s$, and consequently the $\alpha_k$'s are all actually functions of $z$. Their variations along the propagation axis is then taken into account, unlike most of the reported models in which effective parameters are used \cite{Agrawal1988,Su2006,Mork2005}.

\subsection{Small RF signal formulation} \label{ss1}

In order to solve the propagation equation (\ref{eq2}), one first has to determine the expressions of the gain harmonic components $g_k$'s, as functions of the harmonic components of the optical intensity  $M_k$'s, of $\bar{g}$, and of $I_s$ and $\tau_s$. 
Under small RF signal approximation, i.e. considering a small modulation index of the optical carrier, one can assume that $|M_{k-1}|\gg|M_{k}|$ for $k>0$ (and $|M_{k-1}|\ll|M_{k}|$ for $k<0$). It can also be noticed that the same relations hold for the $N_k$'s since the carrier density oscillations are induced by the illumination oscillations through gain saturation. On the basis of this assumption, the commonly used expression for the coefficients $g_k$ is \cite{Herrera2003,Mukai1990,ODuill2009}:
\begin{equation}
g_k=\bar{g}\frac{-M_k/I_s}{1+M_0/I_s+jk\Omega\tau_s}\, , \label{eq7}
\end{equation}
where :\\
\begin{equation}
\bar{g}=\hbar\omega(I/qV-\bar{N}/\tau_s)/M_0\, . \label{eq8}
\end{equation}
In other words, this is equivalent to consider as nonzero only the diagonal and center column of the matrix in (\ref{eq6}). However, the quantities to be compared in the inversion of (\ref{eq6}) are not the $|M_k|$'s, but the terms $|M_{k-p}\times g_p|$. Consequently, the hypothesis $|M_{k-1}|\gg|M_{k}|$ leads to neglect only the terms for which $|p|+|k-p|>|k|$, the others being of the same order. Under these conditions, (\ref{eq6}) is reduced into:
\begin{equation}
\begin{picture}(350,120)
\thicklines
\dottedline{5}(145,96)(190,71)
\dottedline{5}(220,54)(263,27)
\thinlines
\drawline(165,96)(192,81)
\drawline(218,42)(243,26)
\put(10,60){$\left( \begin{array}{c}
0\\
\vdots \\
0\\
\hbar\omega \left(\frac{I}{qV}-\frac{\bar{N}}{\tau_s}\right)\\
0\\
\vdots \\
0
\end{array}\right)=
\left( \begin{array}{ccccccc}
\alpha_{-K} & M_{-1} & \hdots & M_{-K} & 0 & & 0 \\
0 & & & \vdots & & & \\
 & &  & M_{-1} & & & \\
 & & 0 & \alpha_0 & 0 & & \\
 & & & M_1 & & & \\
 & & & \vdots & & & 0\\
0 & & 0 & M_{K} & \ldots & M_1 & \alpha_{K} 
\end{array}\right)\times
\left( \begin{array}{c}
g_{-K}\\
\vdots \\
g_{-1}\\
\bar{g}\\
g_1\\
\vdots \\
g_K
\end{array}\right)$}
\end{picture}
\label{eq9}
\end{equation}

The resolution of this equation gives identical expressions for $\bar{g}$ and $g_{\pm1}$ as in (\ref{eq7}) and (\ref{eq8}), indicating no change on the fundamental component of the output modulation, compared to common models. However, for $|k|>1$, additional terms appear, whose importance will be emphasized later on. The expressions of $g_{2}$ and $g_{3}$ are hence:
\begin{eqnarray}
g_2=\bar{g}\left(\frac{-M_2/I_s}{1+M_0/I_s+2j\Omega\tau_s} + \frac{(M_1/I_s)^2}{(1+M_0/I_s+j\Omega\tau_s)(1+M_0/I_s+2j\Omega\tau_s)}\right)\, , \label{eq10}  \\
g_3=\bar{g}\left(\frac{-M_3/I_s}{1+M_0/I_s+3j\Omega\tau_s} 
 +  \frac{M_1M_2/I_s^2}{(1+M_0/I_s+j\Omega\tau_s)(1+M_0/I_s+3j\Omega\tau_s)} \right. \nonumber \\
 + \frac{M_1M_2/I_s^2}{(1+M_0/I_s+2j\Omega\tau_s)(1+M_0/I_s+3j\Omega\tau_s)} \nonumber \\
\left. - \frac{(M_1/I_s)^3}{(1+M_0/I_s+j\Omega\tau_s)(1+M_0/I_s+2j\Omega\tau_s)(1+M_0/I_s+3j\Omega\tau_s)}\right)\, . \label{eq11}
\end{eqnarray}
Once again, $I_s$ and $\tau_s$ in equations (\ref{eq7}) to (\ref{eq11}) are functions of the propagation coordinate $z$, through their dependence on the static carrier density $\bar{N}(z)$. Doing so, the relations between $I_s$, $\tau_s$ and the physical device constants are preserved, which ensures the predicting capability of the model when the operating conditions (optical input power or bias current) are changed. As reportes in \cite{Berger}, a simple measurement of the unsaturated gain as a function of the SOA bias current, gives a first relation between $\bar{g}$ and $\bar{N}/\tau_s$:
\begin{equation}
\bar{g}=\alpha + \beta \frac{\tau_s}{\overline{N}}\, , \label{eq12}
\end{equation}
where $\alpha$ and $\beta$ are empiric coefficients determined experimentally by measuring the SOA's small signal gain. For our simulations, we used $\alpha=5.88\cdot10^3\mathrm{m^{-1}}$ and $\beta=-1.84\cdot10^{37}\mathrm{m^2.s^{-1}}$, corresponding to the COVEGA InGaAsP/InP quantum well SOA available in our laboratory \cite{Berger}. Then, solving the system constituted of (\ref{eq8}) and (\ref{eq12}), we obtain $\bar{g}(z)$ and $\bar{N}/\tau_s(z)$ as functions of $\alpha$, $\beta$, $I$ and the local DC optical intensity $M_0(z)$. Finally, we model the carrier lifetime in our SOA using the well known expression:
\begin{equation}
\frac{1}{\tau_s}=A+B\bar{N}+C\bar{N}^2 \, , \label{eq13}
\end{equation}
where $A$, $B$ and $C$ are the carrier recombination coefficients of the semiconductor structure, and correspond respectively to the non-radiative, spontaneous, and Auger recombination coefficients. Equation (\ref{eq13}) and the expressions of $\bar{g}(z)$ and $\bar{N}/\tau_s(z)$ then enable to derive $\tau_s$ and $I_s$ as functions of $I$, $M_0(z)$ and $A$, $B$ and $C$. \\
According to \cite{Berger}, $A$, $B$ and $C$ are the only adjustment parameters of the model: they are determined by adjusting the simulated and measured fundamental RF transfer functions of the SOA. In the case of the COVEGA SOA we consider in this paper, we obtained 
\mbox{$A= 2\cdot10^{9} \mathrm{s^{-1}}$}, \mbox{$B=1.2\cdot10^{-10} \mathrm{cm^3.s^{-1}}$}, \mbox{$C=1.8\cdot 10^{-31} \mathrm{cm^6.s^{-1}}$}. It is worthwhile to mention that these values were found to be valid for the full range of injected currents or optical input powers, proving the predictive capability of this approach.

\subsection{Large RF signal formulation} \label{ss2}

In the case of a large modulation index, the previous assumption $|M_{k-1}|\gg|M_{k}|$ do no longer apply. Equation (\ref{eq6}) has to be rigorously solved. Moreover, equation (\ref{eq8}) is no longer valid, and consequently, $\bar{g}$, $I_s$ and $\tau_s$ cannot be obtained as in the small signal case. We thus use the following iterative procedure: in a first step, we substitute $\bar{N}/\tau_s$, $I_s$ and $\tau_s$ in (\ref{eq6}) by their small signal values $\bar{N}/\tau_s^{(0)}$, $I_s^{(0)}$ and $\tau_s^{(0)}$ as obtained in section \ref{ss1}. The central matrix of (\ref{eq6}), referred as $D$, can then be inversed. The coefficients of $D^{-1}$ are denoted $(\delta_{k,p})$. Equations (\ref{eq7}) and (\ref{eq8}) then becomes:
\begin{eqnarray}
\bar{g}&=&\hbar\omega(I/qV-\bar{N}/\tau_s)\delta_{0,0}\, , \\ \label{eq14}
g_k&=&\bar{g}\ \frac{\delta_{k,0}}{\delta_{0,0}}\, .\label{eq15}
\end{eqnarray}
Similarly to the small signal case, using equations (\ref{eq12}), (\ref{eq13}) and (\ref{eq14}), we obtain $\bar{N}/\tau_s^{(1)}$, $I_s^{(1)}$ and $\tau_s^{(1)}$ as functions of $I$, $A$, $B$, $C$ and $M_k(z)$. This procedure is repeated until convergence of $\bar{N}/\tau_s^{(n)}$, $I_s^{(n)}$ and $\tau_s^{(n)}$, which typically occurs after a few tens of iterations. 
The propagation equation (\ref{eq2}) can now be expressed in a matrix formulation similarly to \cite{Mukai1990} and numerically solved.
\begin{equation}
\begin{picture}(320,115)
\drawline(111,95)(245,28)
\drawline(105,86)(221,28)
\drawline(105,61)(171,28)
\drawline(136,95)(254,36)
\drawline(192,95)(254,64)
\put(-20,60){$\begin{array}{c}
d\\
\hline dz
\end{array}
\left( \begin{array}{c}
M_{-K}\\
\vdots \\
M_0\\
\vdots \\
M_K
\end{array}\right)=
\left( \begin{array}{ccccccc}
-\gamma_i+\Gamma \bar{g} & \Gamma g_{-1} & \hdots & \Gamma g_{-K} & 0 & & 0 \\
\Gamma g_1 & & & & & & \\
\vdots & & & & & & 0 \\
\Gamma g_K & & & & & & \Gamma g_{-K} \\
0 & & & & & & \vdots \\
 & & & & & & \Gamma g_{-1} \\
0 & & 0 & \Gamma g_K & \ldots & \Gamma g_1 & -\gamma_i+\Gamma \bar{g} 
\end{array}\right)\times
\left( \begin{array}{c}
M_{-K}\\
\vdots \\
M_0\\
\vdots \\
M_K
\end{array}\right)$}
\end{picture}
\label{eq16}
\end{equation}

\subsection{Comparison with reported models} \label{ss3}
\begin{figure}[h!]
\centering
\subfigure[\label{fig1}]{\includegraphics[height=4.2cm]{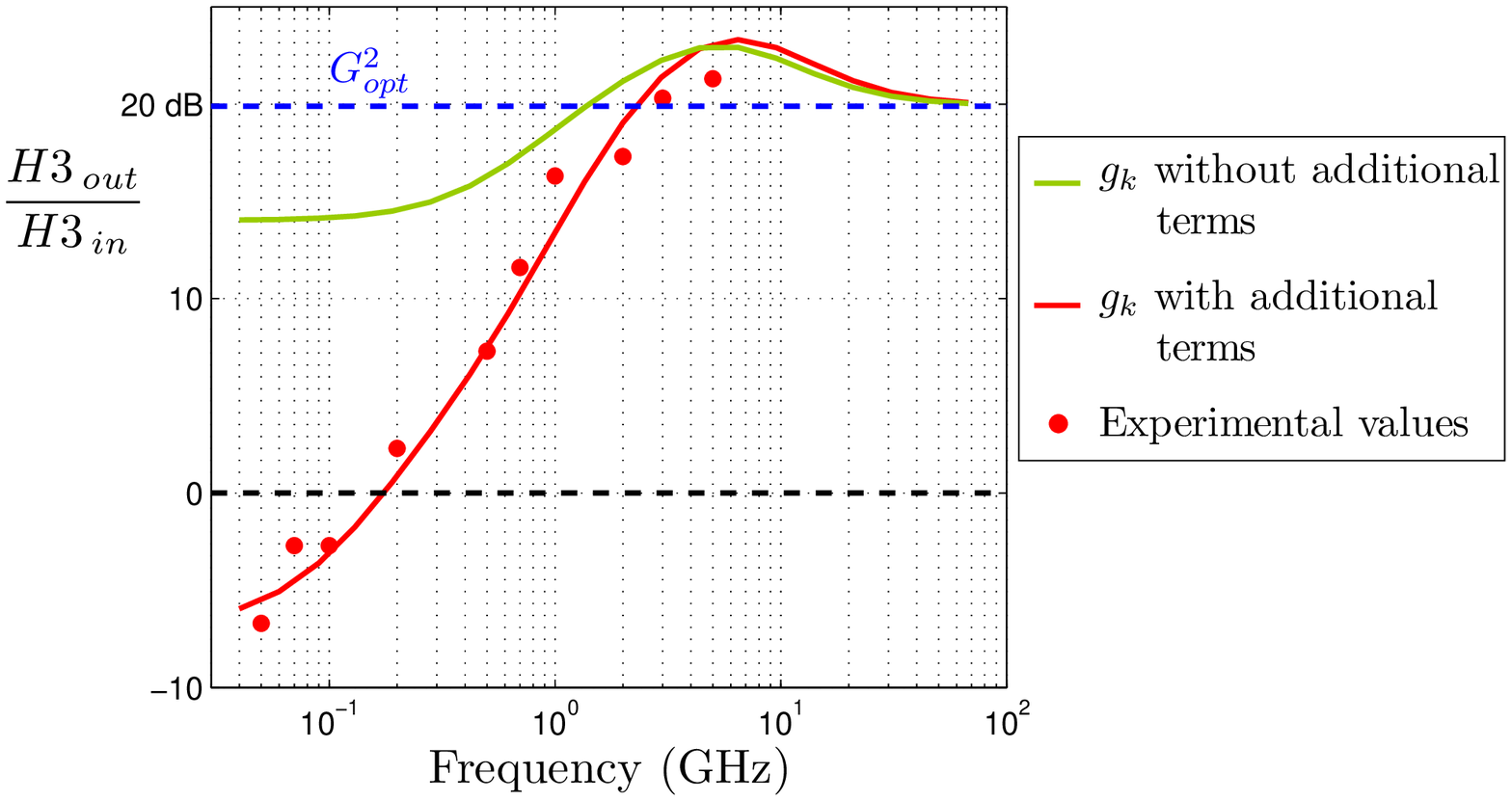}} 
\subfigure[\label{fig2}]{\includegraphics[height=4.2cm]{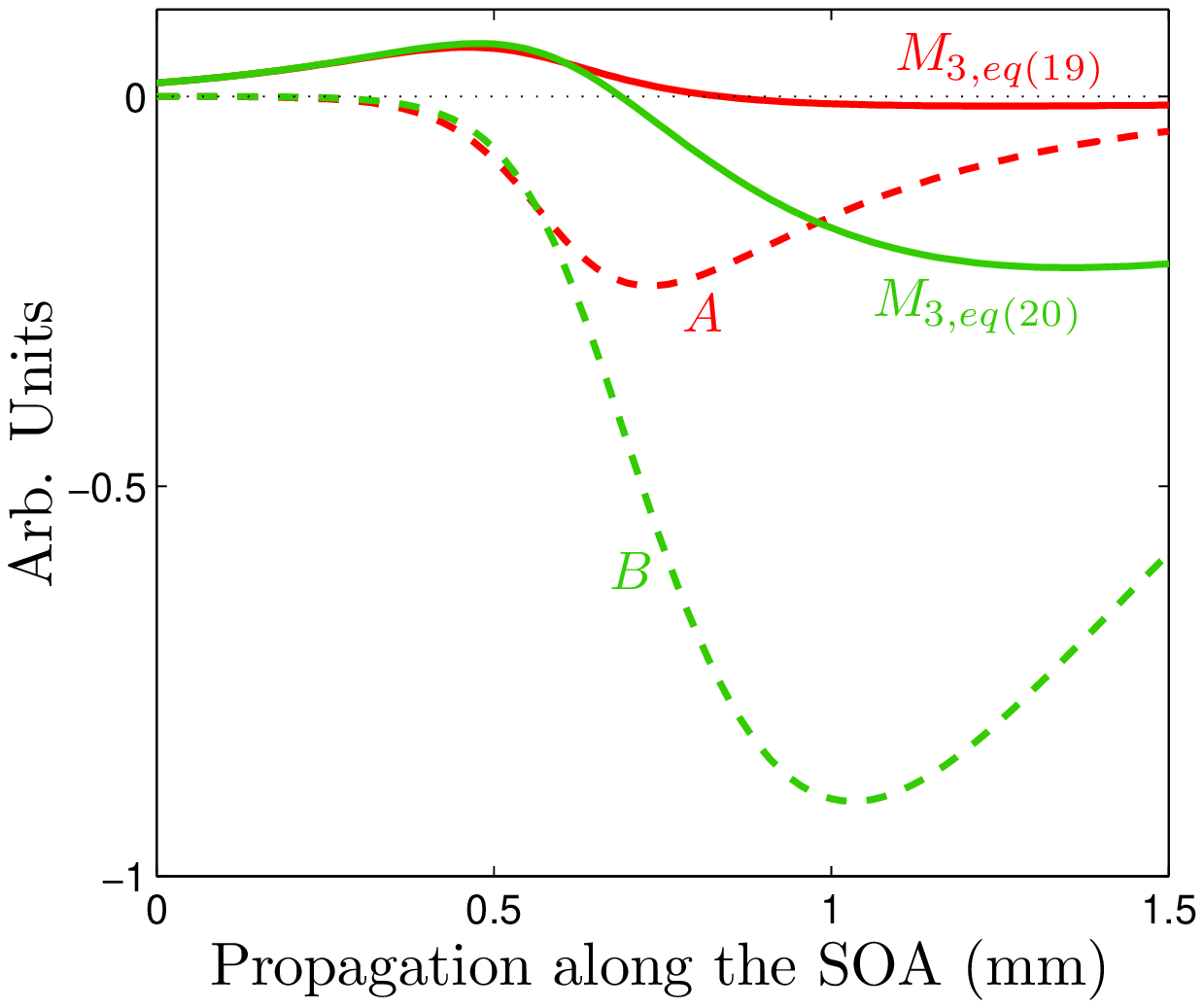}} 
   \caption{(a) Third harmonic generation as a function of the RF frequency. Green solid line: common expressions of $g_k$ (equations (\ref{eq7}) and (\ref{eq8})) are used. Red solid line: use of the rigorous expressions of $g_k$ (equations (\ref{eq10}) and (\ref{eq11})). The red circles represent experimental measurements. (b) Asymptotic case when $\Omega \rightarrow 0$: Evolution of terms $A$ and $B$ along the SOA, and third harmonic level calculated according to (\ref{eq19}) (red solid line) and (\ref{eq19a}) (green solid line).}
\end{figure}

To demonstrate the importance of the additional terms in the right hand side of the expressions of $g_2$ and $g_3$ in equations (\ref{eq10}) and (\ref{eq11}), we calculated the third harmonic power $H_3$ at the output of a $1.5$mm-long SOA. 
The optical carrier is modulated using a Mach-Zehnder modulator. The initial conditions for the resolution of equation (\ref{eq16}) are hence \cite{Willner2003}:
\begin{equation}
\left\{\begin{array}{l}
M_{0,in}=P_{in}\\
M_{1,in}=-P_{in}\times J_1(m)\\
M_{2,in}=0\\
M_{3,in}=P_{in}\times J_3(m)\\
M_{-k,in}=M_{k,in} 
\end{array}\right. \, ,\label{eq17}
\end{equation}
where $m$ is the modulation index, 
and $J_k$ denotes the $k^{th}$ order Bessel function of the first kind. For this example, we chose a modulation index of $0.01$, which ensures the validity of the small RF signal hypothesis.
\begin{table}[!h]
\centering
\begin{tabular}{c|c}
SOA~parameter & Value \\
\hline \hline
Facet to facet linear gain & 37.6dB @I=500mA \\
\hline
Saturation power & 15dBm @I=500mA \\
\hline
Linear losses & 13.6cm$^{-1}$ \\
\hline
Coupling losses & 3.3dB/interface \\
\hline
Amplifier length & 1.5mm
\end{tabular}
\caption{List of studied SOA parameters}
\label{table1}
\end{table}
\\On the other hand, we measured experimentally the third harmonic level at the input and at the output of a commercial SOA, whose main physical parameters are listed in Table \ref{table1}. These parameters were also used in the simulations. It is worth mentioning that the saturation power given in the table is the measured output power for which the unsaturated optical gain is reduced by 3dB. Therefore it does not correspond to $Is$ in the model, which is a local parameter, varying along the propagation axis. The optical input power is 10dBm, for which the optical gain is strongly saturated, and which leads to the maximum RF phase shift that we obtained with this device \cite{Berger}. Equation (\ref{eq16}) is solved using the expression of the $g_k$'s given first by the common equations (\ref{eq7}) and (\ref{eq8}), and then by equations (\ref{eq10}) and (\ref{eq11}) containing the additional terms. Figure \ref{fig1} shows both experimental (red circles) and theoretical (solid lines) third harmonic photodetected power, normalized to the third harmonic power at the input of the SOA. The theoretical curves are evaluated according to:
\begin{equation}
H_3=2R\eta_{ph}^2|M_{3,out}\times S|^2 \label{eq18}
\end{equation}
 where R and $\eta_{ph}$ are respectively the photodiode resistive load ($50\Omega$) and efficiency (equal to 0.8). $S$ denotes the SOA modal area.
 
As can be seen on figure \ref{fig1}, for a modulation frequency $\Omega$ above $4GHz$, the two models are equivalent, and the third harmonic power naturally tends to the value it would have if no CPO effects were present (i.e. $M_{3,out}$ is equal to $M_{3,in}$ multiplied by the optical gain $G_{opt}$). However, at low frequencies, the difference between the two models reaches 20dB. The experimental measurement shows a very good agreement with our model including the additional terms, which confirms the validity of our approach. The large discrepancy with common models can be qualitatively understood considering the asymptotic case where $\Omega$ tends to zero. Using equations (\ref{eq10}) and (\ref{eq11}) in the propagation equation (\ref{eq16}), one obtains for the third harmonic term, with $\Omega\tau_s\ll1$:
\begin{equation}
\frac{dM_3}{dz}=-\gamma_i M_3+\bar{g}\frac{I_s}{M_0+I_s}\left[M_3\underbrace{-2\frac{M_1M_2}{M_0+I_s}+\frac{M_1^3}{(M_0+I_s)^2}}_{A}\right]\, , \label{eq19}
\end{equation}
whereas, using the standard models based on equation (\ref{eq7}), one has:
\begin{equation}
\frac{dM_3}{dz}=-\gamma_i M_3+\bar{g}\frac{I_s}{M_0+I_s}\left[M_3\underbrace{-2\frac{M_1M_2}{I_s}}_{B}\right]\, . \label{eq19a}
\end{equation}
Basically, the right-hand side of these two equations consists in three terms. The first two respectively describe the linear losses and amplification of $M_3$. The third one, respectively denoted $A$ and $B$ for equations (\ref{eq19}) and (\ref{eq19a}) describe the energy transfer from $M_1$ and $M_2$ to $M_3$ through CPO process. Both $A$ and $B$ have a negative sign, which is consistent with the CPO's oscillations being in antiphase with the illumination oscillations. In figure \ref{fig2} are plotted the evolution along the propagation axis inside the SOA of terms $A$ and $B$, and of $M_3$, calculated according to equation (\ref{eq19}) in red, and according to equation (\ref{eq19a}) in green. These results indicate that $|B|$ is much larger than $|A|$, and hence that conventional models overestimates the third harmonic generation term due to CPO for the low frequency part of the spectrum. Consequently, CPO contribution largely dominates the amplification term, and leads to a relatively high output level of third harmonic. On the other hand, when the gain harmonics are rigorously derived, one ends up with the term $A$, with a smaller magnitude which balances the amplification process. The resulting output $H_3$ level is therefore significantly reduced, and can even be under the input level, if the optical gain is saturated enough such as the overall amplification/generation terms are below the linear losses attenuation.\\  
This shows the importance of considering accurate expressions for $g_k$ even in the small signal situation, and especially for the low frequency part of the microwave spectrum. When $\Omega$ increases, since the  $2^{nd}$ and $3^{rd}$ terms of (\ref{eq11}) respectively evolve as $\Omega^{-1}$ and $\Omega^{-2}$, then equation (\ref{eq11}) progressively tends to the common expression given by equation (\ref{eq7}). This can be seen on figure \ref{fig1}, where the two curves finally coincide for $\Omega>4GHz$. Consequently, the present analysis does not question the results obtained for instance in \cite{ODuill2009}, where harmonics calculations are carried out at a fixed frequency of $10GHz$. However, as confirmed by experiments, a rigorous calculation is mandatory when the modulation frequency lies in the spectral region $1-3\, \mathrm{GHz}$ where most of ground radars operate.  
  
\section{Intermodulation distortion}\label{s3}
\begin{figure}[h!]
  \centering
  \includegraphics[width=14cm]{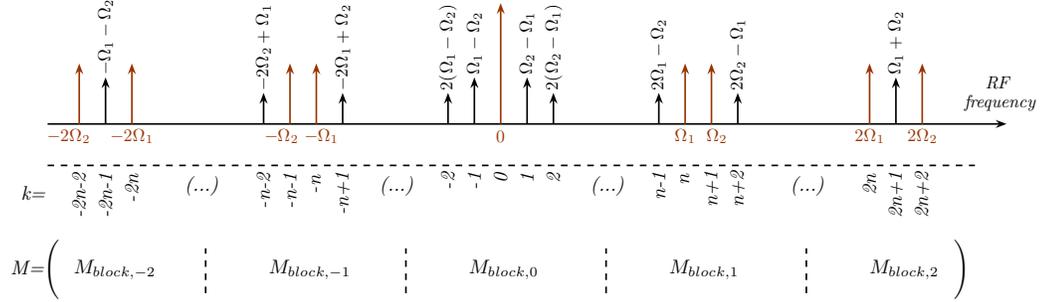}
  \caption{Set of significant spectral components of $|E|^2$, $N$ and $g$, and associated index $k$ in their Fourier decompositions. $n$ is defined such as $\Omega_1=n\delta\Omega$.}
  \label{fig3}
\end{figure}

Intermodulation distortion ($\mathrm{IMD}$) calculation is slightly different from what has been discussed in the above section. Indeed, the number of mixing terms that must be taken into account is significantly higher. For radar applications a typical situation where the $\mathrm{IMD}$ plays a crucial role is that of a radar emitting at a RF frequency $\Omega_1$, and facing a jammer emitting at  $\Omega_2$, close to $\Omega_1$. Both $\Omega_1$ and $\Omega_2$ are collected by the antenna and transferred to the optical carrier through a single electro-optic modulator. The point is then to determine the nonlinear frequency mixing due to the coherent population oscillations (CPO) inside the SOA. In particular, the mixing products at frequencies $\Omega_2-\Omega_1$ (or $\Omega_1-\Omega_2$) and $2\Omega_2-\Omega_1$ (or $2\Omega_1-\Omega_2$) ---~respectively called second ($\mathrm{IMD_2}$) and third ($\mathrm{IMD_3}$) order intermodulation distortions~--- have to be evaluated at the output of the SOA.\\
The main difference with harmonic calculation is that the optical intensity, and hence the SOA carrier density $N$, and the SOA gain $g$ are no longer time-periodic functions of period $\Omega$, but of period $\delta\Omega=\Omega_2-\Omega_1$. Then, using a Fourier decomposition of $|E|^2$, $N$, and $g$, one has:
\begin{eqnarray}
|E(z,t)|^2&=&\sum_{k=-\infty}^{+\infty} M_k(z) e^{-ik\delta\Omega t}\, , \label{eq20}\\
N(z,t)&=&\bar{N}(z) + \sum_{\substack{k=-\infty \\ k\neq0}}^{+\infty} N_k(z) e^{-ik\delta\Omega t}\, , \label{eq21}\\
g(z,t)&=&\bar{g}(z) + \sum_{\substack{k=-\infty \\ k\neq0}}^{+\infty} g_k(z) e^{-ik\delta\Omega t}\, . \label{eq22}
\end{eqnarray}
We consider a typical radar frequency $\Omega_1$ of 10GHz, and a frequency spacing $\delta\Omega$ of 10MHz. Here, for intermodulation distortion calculation, we assume that only the spectral components at $\Omega_{1,2}$, $2\Omega_{1,2}$, and all their first order mixing products significantly contribute to the generation of $\mathrm{IMD_2}$ and $\mathrm{IMD_3}$, as illustrated in figure \ref{fig3}. The $M_k$'s and the $g_k$'s  are then reduced in 19 elements vectors, and gathered into blocks, denoted $M_{block,j}$ and $g_{block,j}$. The $j^{th}$ block contains the mixing products with frequencies close to $j\times \Omega_1$. According to the notation of figure \ref{fig3}, the carrier rate equation (\ref{eq1}) and the propagation equation (\ref{eq2}) can respectively be written as:
\begin{equation}
\left( \begin{array}{c}
0\\
0\\
\hbar\omega \left(I/qV-\bar{N}/\tau_s\right)\\
0\\
0
\end{array}\right)=
\left( \begin{array}{ccccc}
D_{-2,-2} & D_{-2,-1} & D_{-2,0} & 0 & 0 \\
D_{-1,-2} & D_{-1,-1} & D_{-1,0} & D_{-1,1} & 0 \\
D_{0,-2} & D_{0,-1} & D_{0,0} & D_{0,1} & D_{0,2} \\
0 & D_{1,-1} & D_{1,0} & D_{1,1} & D_{1,2} \\
0 & 0 & D_{2,0} & D_{2,1} & D_{2,2} \\
\end{array}\right)\times
\left( \begin{array}{c}
g_{block,-2} \\
g_{block,-1} \\
g_{block,0} \\
g_{block,1} \\
g_{block,2} \\
\end{array}\right)\, ,\label{eq23}
\end{equation}
~
\begin{equation}
\frac{d}{dz}\left( \begin{array}{c}
M_{block,-2} \\
M_{block,-1} \\
M_{block,0} \\
M_{block,1} \\
M_{block,2}
\end{array}\right)=
\left( \begin{array}{ccccc}
H_{-2,-2} & H_{-2,-1} & H_{-2,0} & 0 & 0 \\
H_{-1,-2} & H_{-1,-1} & H_{-1,0} & H_{-1,1} & 0 \\
H_{0,-2} & H_{0,-1} & H_{0,0} & H_{0,1} & H_{0,2} \\
0 & H_{1,-1} & H_{1,0} & H_{1,1} & H_{1,2} \\
0 & 0 & H_{2,0} & H_{2,1} & H_{2,2}
\end{array}\right)\times
\left( \begin{array}{c}
M_{block,-2} \\
M_{block,-1} \\
M_{block,0} \\
M_{block,1} \\
M_{block,2} \\
\end{array}\right)\, ,\label{eq24}
\end{equation}
where $D_{i,j}$ and $H_{i,j}$ are sub-blocks of the complete matrices of equations (\ref{eq6}) and (\ref{eq16}) respectively. They  describe the mixing of the spectral components around the $i^{th}$ and the $j^{th}$ harmonics. From now on, the full procedure described in section \ref{s2} can be applied in the same iterative way to determine the $g_k$'s, $I_s$ and $\tau_s$, and to finally numerically solve the equation (\ref{eq24}). Similarly to equation (\ref{eq18}), the photodetected RF power at $2\Omega_2-\Omega_1$ is then calculated through:
\begin{equation}
\mathrm{IMD_3}=2R\eta_{ph}^2|M_{2\Omega_2-\Omega_1}^{out}\times S|^2\, . \label{eq25}
\end{equation}

\section{Simulation results}\label{s4}
\begin{figure}[h!]
\centering
\subfigure[\label{fig4}]{\includegraphics[height=4.5cm]{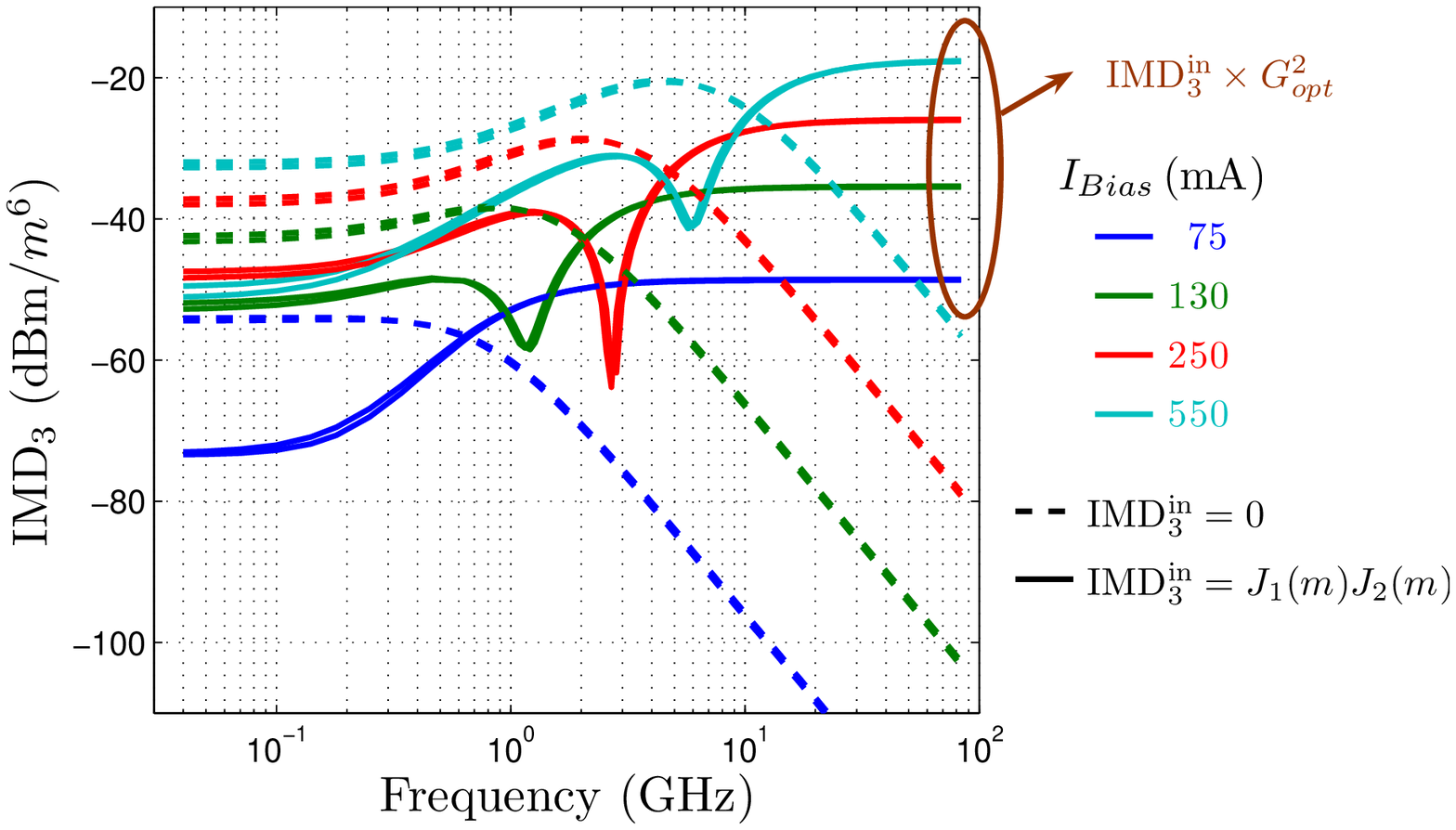}}
\subfigure[\label{fig5}]{\includegraphics[height=4.5cm]{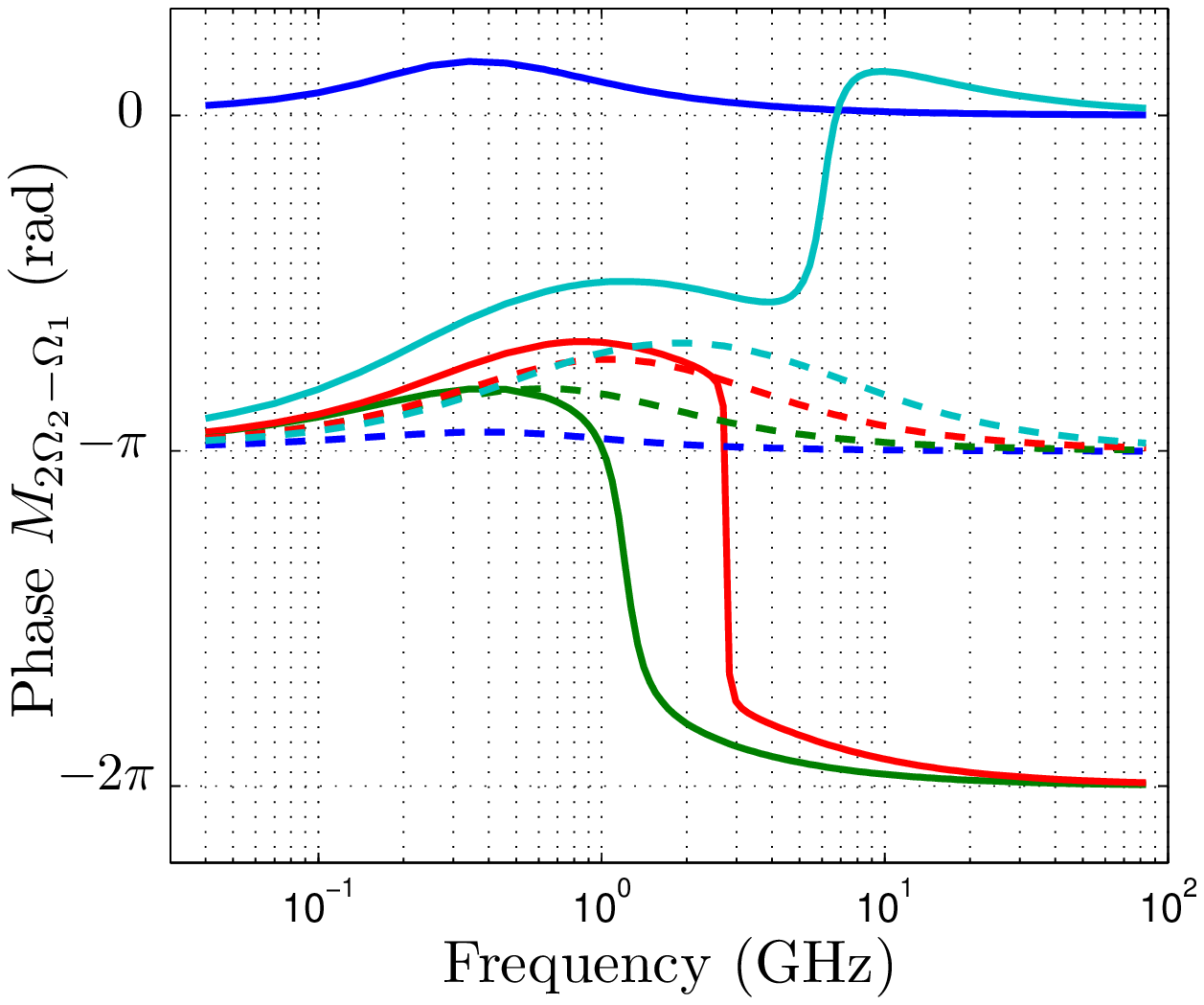}}
   \caption{(a) Computed third order intermodulation power $\mathrm{IMD_3}$, normalized to $m^6$, as a function of the modulation frequency, for various injected currents. (b) Corresponding phase of the beat-note at $2\Omega_2-\Omega_1$. Dashed lines: case of a perfectly linear modulator ($\mathrm{IMD_3^{in}}=0$). Solid lines: actual Mach-Zehnder modulator ($\mathrm{IMD_3^{in}}\neq0$).}
\end{figure}

In the general situation depicted in section \ref{s3}, the optical intensity at the output of a chirp-free Mach-Zehnder modulator, that is, at the input of the SOA, will be of the following form \cite{Willner2003}:
\begin{equation}
I_{opt,in}=I_0\bigg[1+cos\Big(m \big(cos(\Omega_1t)+cos(\Omega_2t)\big)+\phi\Big)\bigg]\, , \label{eq26}
\end{equation}
where $m$ is the modulation index, and $\phi$ is the modulator phase bias. When developed into first kind Bessel functions,  and according to the notation of figure \ref{fig3}, the corresponding input distribution of the $M_k$'s, at quadrature bias ($\phi=\pi/2$), is:
\begin{equation}
M_{block,0}=\left|\begin{array}{c}
0\\
0\\
1\\
0\\
0
\end{array}\right.;\ \ 
M_{block,1}=\left|\begin{array}{c}
J_1(m)J_2(m)\\
-J_0(m)J_1(m)\\
-J_0(m)J_1(m)\\
J_1(m)J_2(m)
\end{array}\right.;\ \ 
M_{block,2}=\left|\begin{array}{c}
0\\
0\\
0
\end{array}\right.;\ \ 
M_{block,-j}=M_{block,j}. \label{eq27}
\end{equation}
In particular, one can notice that the optical intensity at the input of the SOA contains a term at $2\Omega_{1,2}-\Omega_{2,1}$, denoted $\mathrm{IMD_3^{in}}$, and equal to $J_1(m)J_2(m)$. In a general way, when considering the propagation equation in its matricial form (eqs. (\ref{eq16}) and (\ref{eq24})), the propagation of the Fourier compounds of an optically carried microwave signal into the SOA can be seen as resulting from an amplification process (the diagonal terms of the matrix) and a generation process by frequency mixing through coherent population oscillations (non-diagonal terms of the matrix). In order to better understand the combination of these two effects in the case of $\mathrm{IMD_3}$ propagation, we conducted our simulations considering two initial conditions, with and without $\mathrm{IMD_3^{in}}$. The latest case corresponds to the use of a perfectly linear modulator, or as in \cite{Seo2003} for Radio-over-Fiber applications, where modulations at $\Omega_1$ and $\Omega_2$ are produced by two distinct modulators on two incoherent optical carriers. An alternative way to avoid $\mathrm{IMD_3^{in}}$ would be to use a Single-Side-Band modulator, as proposed in \cite{Breuil2004,Alouini2001}. In a tunable delay line based on slow-light effects in a SOA, the optical group delay can be typically tuned either by varying the optical input power, or the SOA bias current. Although both situations can be easily simulated, in this paper, we only present the latter one, using the bias current, which is the most suitable for an implementation in a real radar system. Figure \ref{fig4}  represents the simulated $\mathrm{IMD_{3}}$, calculated according to equation (\ref{eq25}), as a function of the modulation frequency $\Omega_1$, and for a set of SOA injected currents, ranging from $75\, \mathrm{mA}$, slightly above the transparency ($50\, \mathrm{mA}$), up to the maximal current of $550\, \mathrm{mA}$. $\mathrm{IMD_{3}}$ is known to scale with $(P_{RF}^{in})^3$, where $P_{RF}^{in}$ the RF power applied to the Mach-Zehnder modulator. Moreover, $P_{RF}^{in}$ is proportional to $m^2$. Thus, in order to get rid of the dependence on the modulation index, we chose to normalize $\mathrm{IMD_3}$ by $m^6$. In figure \ref{fig4} the simulation results are plotted for $m=0.01$ and $m=0.5$. As expected, the curves are perfectly coincident.  

The expected low-pass filter behavior of the SOA is observed on figure \ref{fig4}, when no beat-note at $2\Omega_2-\Omega_1$ is injected into the SOA (dashed lines) \cite{Herrera2003}. The situation is completely different when the nonlinearities of the modulator used to transfer the RF signal to the optical carrier are taken into account in the simulations (solid lines). First, as for the third harmonic generation (see section \ref{ss3}), the asymptotic values for the high frequency part of the spectrum naturally corresponds to $\mathrm{IMD_3^{in}}$ multiplied by $G_{opt}^2$, for the same reason as the low-pass behavior mentioned above, i.e. that CPO effects are roughly restricted to frequencies below $1/\tau_s$. Conversely, for the low frequency part of the spectrum, the intermodulation products can be seen as the result of the combination of two effects, namely the amplification of the incident beat-note at $2\Omega_2-\Omega_1$, and the generation of a $2\Omega_2-\Omega_1$ mixing term due to the gain modulation induced by the CPO effects inside the SOA. The major difference between these two effects is that the amplified beat-note is roughly in phase with the incident one, whereas for the CPO effects, due to the $\pi$-phase shift between the intensity and gain modulations, the phase of the CPO-induced beat-note at $2\Omega_2-\Omega_1$ is also $\pi$-shifted with respect to the incident beat-note. To better understand the results of figure \ref{fig4}, and particularly the dip observed in the $\mathrm{IMD_3}$ curves, we also computed the evolution of the phase of the beat-note term $M_{2\Omega_2-\Omega_1}$ versus modulation frequency (see figure \ref{fig5}). At low frequencies below $1/\tau_s$, CPO effects dominates, as indicated by the phase curves distribution around $-\pi$, whereas at high frequencies, the pure amplification dominates with phases distributed around $0$ and $-2\pi$. The dip observed in the $\mathrm{IMD_3}$ curves around $1/\tau_s$ therefore corresponds to a transition between these two regimes, where the two contributions tend to cancel each other as they are of opposite signs. This is also confirmed by the $\pi$-phase shift at the dip frequency that is seen on figure \ref{fig5}. It can also be noticed that when the SOA gain gets weaker, as for the solid blue curve ($I_{bias}=75\, \mathrm{mA}$), then the amplification process always dominates over the CPO effect, even if they are probably quite balanced at low frequencies, explaining the low level of $\mathrm{IMD_3}$ in this frequency domain.        

\section{Spurious-Free Dynamic Range}\label{s5}
\begin{figure}[h!]
  \centering
  \includegraphics[height=4.5cm]{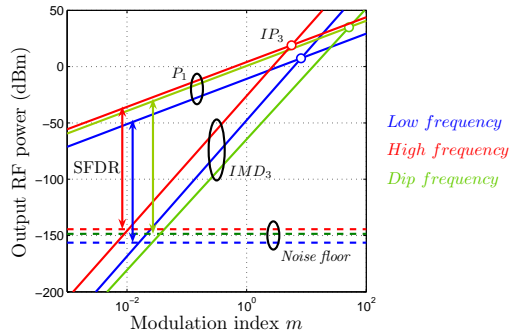}
  \caption{SFDR determination for $I_{bias}=250\, \mathrm{mA}$, $P_{in}=10\, \mathrm{mW}$, and $\mathrm{IMD_3^{in}}=J_1(m)J_2(m)$. In blue: low modulation frequency ($\ll 1/\tau_s$); in red: high modulation frequency ($\gg 1/\tau_s$); in green: frequency around the $\mathrm{IMD_3}$ dip ($\approx 1/\tau_s$).}
  \label{fig6}
\end{figure}

The spurious-free dynamic range ($\mathrm{SFDR}$) is a key parameter to evaluate the performance of a microwave photonic link in a radar system. It is determined by the noise floor $P_{Noise}$, and the third order intercept point $\mathrm{IP_3}$. This point corresponds to the extrapolated fundamental RF output power $P_1$ such as $P_1=\mathrm{IMD_3}$. The noise floor is defined as the electrical noise power contained in a $1\, \mathrm{Hz}$ electrical analysis bandwidth. The values of $P_{Noise}$ we used in this paper were both measured and theoretically calculated for our SOA as a function of the bias current and the RF frequency \cite{BergerSFL}. Finally, the SFDR is obtained according to \cite{Zmuda1994}: 
\begin{equation}
SFDR=\left(\frac{IP_3}{P_{noise}}\right)^{2/3}\ , \label{eq28}
\end{equation}
and corresponds to the situation where the spurious (here the $\mathrm{IMD_3}$) equals the noise floor, as illustrated in \ref{fig6}.

\begin{figure}[h!]
\centering
\subfigure[\label{fig7}]{\includegraphics[height=4.5cm]{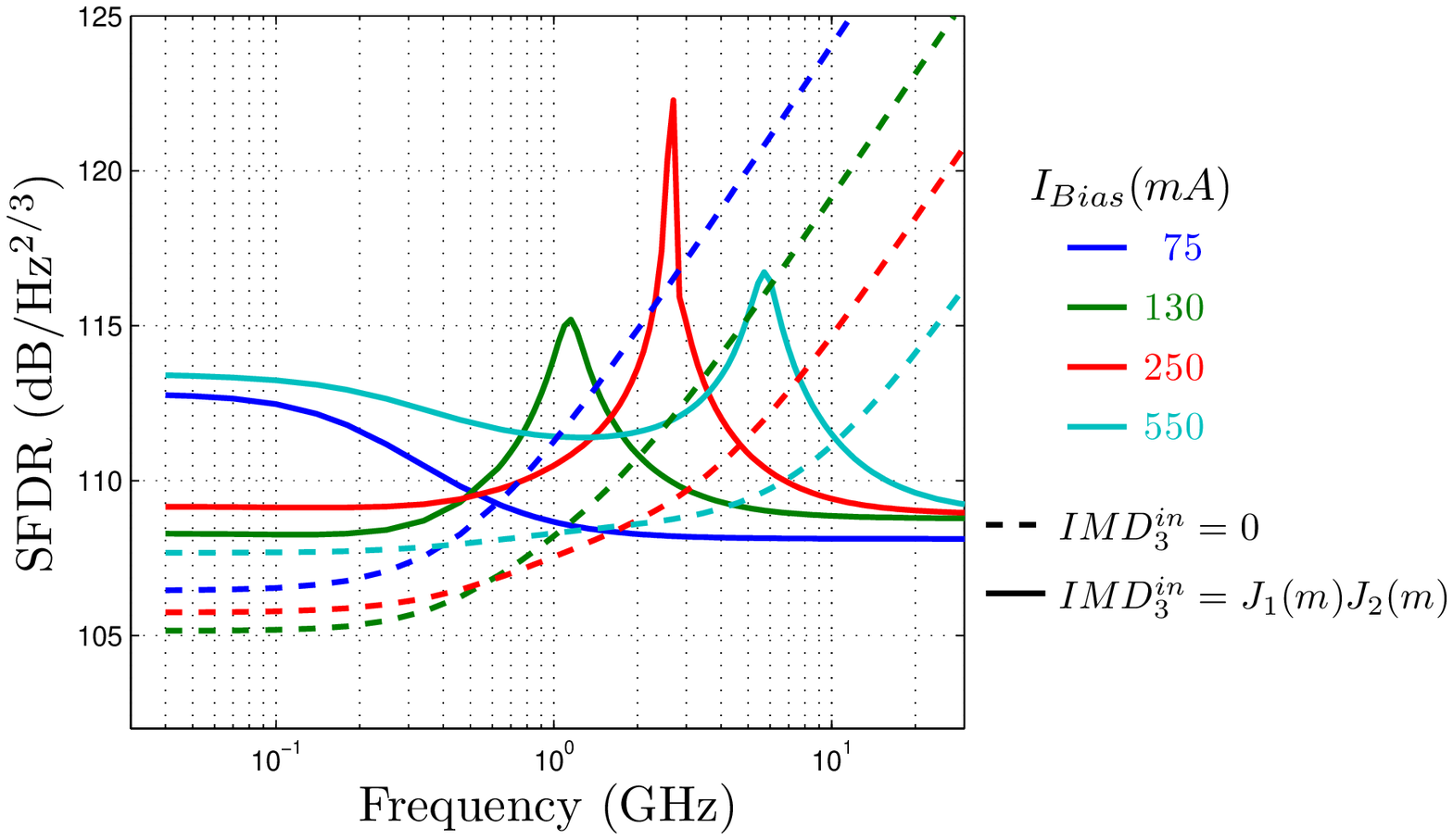}}
\subfigure[\label{fig8}]{\includegraphics[height=4.5cm]{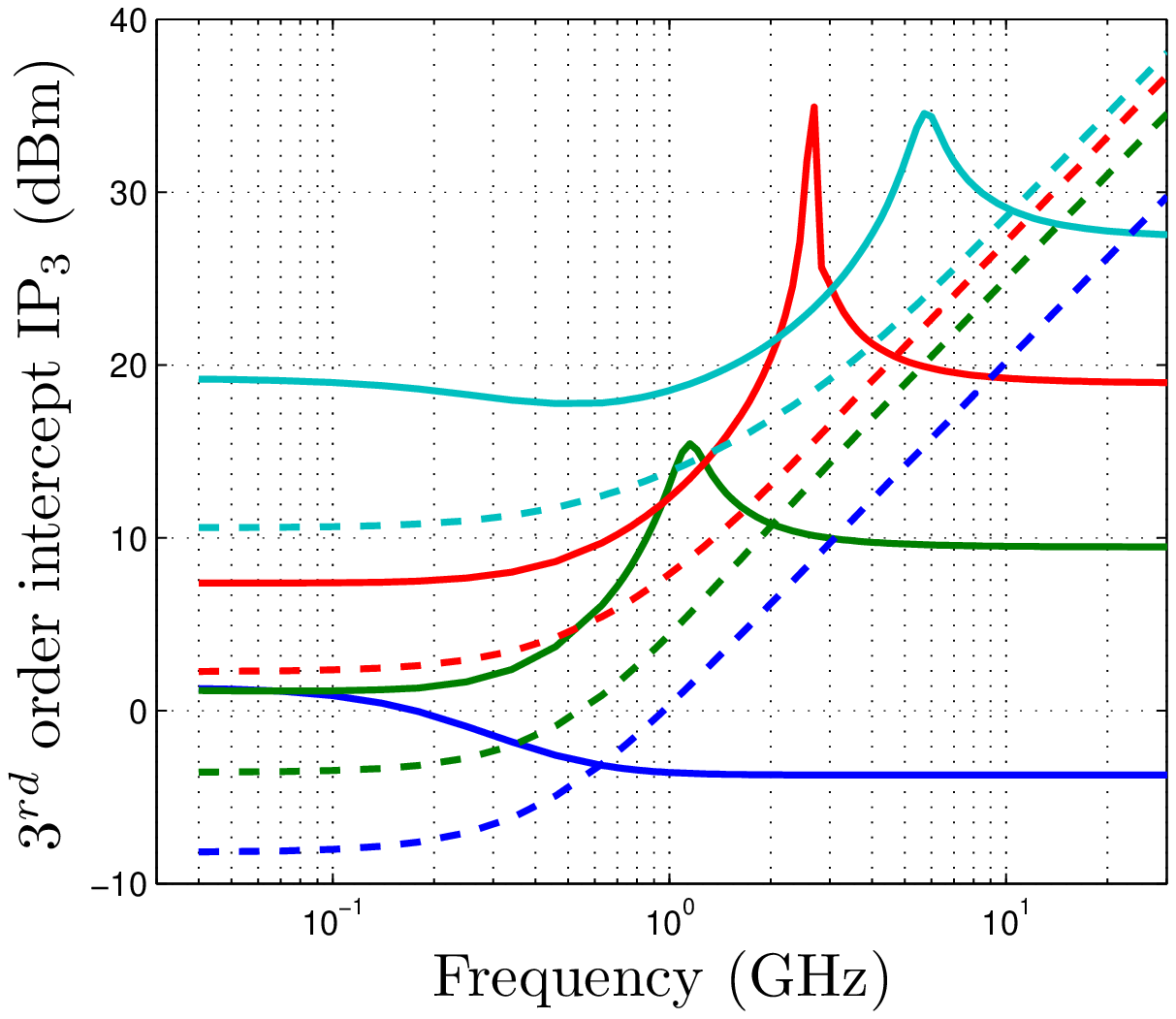}}
   \caption{(SFDR (a) and $3^{rd}$ order intercept point $\mathrm{IP_3}$ (b), as a function of the modulation frequency, for various injected currents, and in the two configurations of a perfectly linear modulation and of a realistic Mach-Zehnder modulator (respectively in dotted and solid lines).}
\end{figure}

We calculated both the SFDR and the third order intercept point $\mathrm{IP_3}$ in the modulation frequency range from $40\, \mathrm{kHz}$ up to $30\, \mathrm{GHz}$, and for various injected currents. Once again, we compared the situation of a perfectly linear modulation, namely without intermodulation products at the input of a SOA, with the case of an actual Mach-Zehnder modulation, with intermodulation products given by the Bessel functions expansion. These two situations are represented in figures \ref{fig7} and \ref{fig8} respectively in dotted and solid lines.\\

A first conclusion is that in both cases, over the full spectral range of interest for slow light applications (between $0.1$ and $10\, \mathrm{GHz}$), and for all considered bias currents, the SFDR of the simulated link scales from $105$ to $125\, \mathrm{dB/Hz^{2/3}}$. These values reasonably match radar system requirements. Moreover, for modulation frequencies below or in the range of $1/\tau_s$, the presence of an initial $\mathrm{IMD_3}$ at the input of the SOA can even improve the SFDR of the link. This phenomenon is widely known among the analog optical transmission community, where the nonlinear gain transfer function of a SOA is used to linearize a Mach-Zehnder modulator and hence reduce the nonlinear distortion effects \cite{Jeon2002,Brock1996}. However, our theoretical analysis of this effect in the framework of microwave photonics, and its interpretation by the input $\mathrm{IMD_3}$ being in antiphase with the CPO gain gratings (see section \ref{s4}), cast new light on this effect, and gives it different perspectives when slow light applications are considered. Moreover, it is also important to notice that the frequency domain where a dip is observed in the $\mathrm{IMD_3}$, and consequently a peak in the SFDR, roughly corresponds to the domain where slow-light effects are the most efficient in the SOA, i.e. where the CPO-induced phase shift is maximal. This last point could be of importance from the operational point of view. 

\section{Conclusion}
We reported in this paper a generalized and predictive model that enables to derive the high order harmonic generation and the intermodulation products inside a SOA. This model takes into account the variations of the gain saturation parameters along the propagation axis into the SOA, and uses a rigorous expression of the gain oscillation harmonics. We also showed that provided a basic small signal gain measurement is conducted, this model only relies on adjustment parameters which are material constants, ensuring the predictive capability of the model. This enabled us to derive the SFDR of a SOA-based microwave photonic link for various experimental conditions. In particular, we investigated the cases where either a perfectly linear or an actual modulator is used at the input of the link. We came to the conclusion that the nonlinear distortion induced by the input modulator could even be favorable to the spurious-free dynamic range of the link, especially in the spectral range where slow light induced phase shifts are the most important. Within the field of analog optical transmission, the usual interpretation of this linear dynamic range improvement is the compensation of the nonlinearities of the sinusoidal transfer function of the Mach-Zehnder modulator by the nonlinearities of the gain transfer function of the SOA. However, by considering slow light applications, we proposed a more in-depth interpretation based on the CPOs in the semiconductor amplifier and including RF frequency dependence, which we believe is therefore more powerful and realistic than the usual one. Finally, the results obtained in this work are encouraging since they show that the nonlinearities brought by a commercial SOA, when operated in the slow light regime, are already acceptable for the less demanding radar applications. 

\section*{Acknowledgments}
 The authors acknowledge the partial support from the GOSPEL European project and from the French ``D\'el\'egation G\'en\'erale pour l'Armement''. The authors would also like to acknowledge Reynald Boula-Picard for early discussions.
\end{document}